\documentclass[letterpaper, 10 pt, conference]{ieeeconf}  % Comment this line out
\usepackage[pdftex,pdfauthor={Quindlen},pdftitle={Closed-Loop Statistical Verification of Stochastic Nonlinear Systems Subject to Parametric Uncertainties}]{hyperref}
\hypersetup{colorlinks,linkcolor={green!50!black},citecolor={green!50!black},urlcolor={blue!80!black}}
\makeatletter \let\NAT@parse\undefined \makeatother

\usepackage{amsmath}
\usepackage[sort,compress]{cite}
\usepackage{graphicx}
\usepackage{epstopdf}
\usepackage{subfigure}
\usepackage{units}
\usepackage{rotating}
\usepackage{graphicx,paralist}
\usepackage{wrapfig}
\usepackage{nomencl}
\usepackage{amssymb}

\renewcommand{\vec}[1]{\boldsymbol{\mathbf{#1}}} %Easier to bold greek letters?

\newcommand{\degrees}{\ensuremath{^\circ}}

\newcommand{\fig}[1] {Figure~\ref{#1}}

\newcommand{\vtheta}{\vec{\theta}}

\usepackage{theorem,balance}
\usepackage{algorithm}%http://rtlab.csie.ntu.edu.tw/resources/Latex/algorithms.pdf
\usepackage{algorithmic}
\theoremstyle{plain} \theorembodyfont{\upshape}

\newtheorem{definition}{\indent Definition}

\newtheorem{assumption}{\indent Assumption}

\usepackage[usenames]{color}
\usepackage[svgnames]{xcolor}
\definecolor{DarkGreen}{rgb}{0,0.5,0}
\definecolor{DarkRed}{rgb}{0.75,0,0}
\usepackage{balance}
\usepackage[font=small]{caption}
\usepackage[normalem]{ulem}
\usepackage[capitalize]{cleveref}
\crefformat{equation}{(#2#1#3)}
\Crefformat{equation}{Equation~(#2#1#3)}
\Crefname{equation}{Equation}{Equations}
\IEEEoverridecommandlockouts % This command is only needed if you want to  use the \thanks command
\renewcommand{\hat}{\widehat}
\renewcommand{\tilde}{\widetilde}
\usepackage[footnote,printonlyused]{acronym}
\usepackage{scrextend}
\deffootnote[0.0em]{0em}{0em}
{\textsuperscript{\thefootnotemark}\,\enskip}

\title{\LARGE \bf
Closed-Loop Statistical Verification of Stochastic Nonlinear Systems Subject to Parametric Uncertainties}

\author{John F. Quindlen$^{1}$%
\thanks{$^{1}$Ph.D. Candidate, Department of Aeronautics and Astronautics, Massachusetts Institute of Technology (MIT).}%
\and Ufuk Topcu$^{2}$\thanks{$^{2}$Assistant Professor, Department of Aerospace Engineering and Engineering Mechanics, University of Texas at Austin}%
\and Girish Chowdhary$^{3}$\thanks{$^{3}$Assistant Professor, Departments of Agricultural \& Biological Engineering and Aerospace Engineering, University of Illinois Urbana-Champaign}%
\and Jonathan P.\ How$^{4}$\thanks{$^{4}$Richard C.\ Maclaurin Professor of Aeronautics and Astronautics, Laboratory for Information and Decision Systems (LIDS), MIT}%
}

\begin{document}
\raggedbottom 
\allowdisplaybreaks
 
\maketitle
\thispagestyle{empty}
\pagestyle{empty}

%%%%%%%%%%%%%%%%%%%%%%%%%%%%%%%%%%%%%%%%%%%%%%%%%%%%%%%%%%%%%%%%%%%%%%%%%%%%%%%%
\begin{abstract}
This paper proposes a statistical verification framework using Gaussian processes (GPs) for simulation-based verification of stochastic nonlinear systems with parametric uncertainties.  Given a small number of stochastic simulations, the proposed framework constructs a GP regression model and predicts the system's performance over the entire set of possible uncertainties.  Included in the framework is a new metric to estimate the confidence in those predictions based on the variance of the GP's cumulative distribution function.  This variance-based metric forms the basis of active sampling algorithms that aim to minimize prediction error through careful selection of simulations.  In three case studies, the new active sampling algorithms demonstrate up to a 35\% improvement in prediction error over other approaches and are able to correctly identify regions with low prediction confidence through the variance metric.
\end{abstract}

%%%%%%%%%%%%%%
\section{Introduction}
Ever increasing demands for high performance necessitates the adoption of advanced, nonlinear control techniques in a wide variety of engineering systems.  The complexity of these controllers, coupled with the increasing complexity of the open-loop systems themselves, complicates verification that the closed-loop system actually satisfies the performance requirements at off-nominal conditions, called parametric uncertainties.  Control system verification analyzes the closed-loop system and attempts to identify at which parametric uncertainties the resulting trajectory fails to meet the requirements.  

Traditional analytical and numerical verification techniques\cite{Clarke99,Prajna06_Automatica} provably verify the system's performance at various uncertainties, but require modeling assumptions that restrict their suitability to complex systems with highly nonlinear dynamics and hybrid controllers.  Recent work\cite{Topcu08_PhD,Majumdar14_CDC,Kapinski14_HSCC} addressed some of these concerns, but are limited by the size of the state and parameter spaces.  Stochastic dynamics further challenge these approaches and restrict their applicability.  For comparison, statistical verification techniques\cite{Zuliani12_HSCC,Clarke11_ATVA} replace provable guarantees with looser probabilistic bounds, but are capable of handling a much broader range of problems with fewer restrictions.  Statistical verification methods rely upon extensive simulations of the system to test the likelihood of requirement satisfaction at various conditions and are well-suited to stochastic systems.  

Recent work developed data-driven statistical verification techniques\cite{Kozarev16_HSCC,Quindlen18_GNC} to efficiently verify deterministic closed-loop systems without relying upon exhaustive simulations.  These data-driven methods construct machine learning models from small sets of simulation trajectories and predict the response at unobserved parameter settings.  The latest work\cite{Quindlen17a_Arxiv} introduced Gaussian process (GP) prediction models and novel active sampling procedures to improve the accuracy of predictions given limited amounts of simulation data.  The deterministic GP-based approach also quantifies the prediction confidence online without the need for external validation datasets and carefully selects simulations to maximize the prediction confidence. 

%\XX{seems like a very long summary of an 8 page paper}
%This paper extends data-driven methods to stochastic verification problems and provides three primary contributions.  First, Section \ref{s:modeling} presents a stochastic statistical verification framework to predict the probability of requirement satisfaction at all possible parametric uncertainties.  This GP-based framework infers the underlying satisfaction probability function from nondeterministic simulations at different parameter settings.  Additionally, Section \ref{s:confidence} introduces a novel validation tool for online identification of regions with low prediction confidence.  Reformulated active sampling procedures in Section \ref{s:active} exploit this new validation metric and iteratively minimize prediction error to create a closed-loop verification process.  Three case studies at the end demonstrate the effectiveness of the paper's contributions for resource-constrained statistical verification.
%\XX{contributions of the paper are:}...

This paper extends data-driven methods to stochastic verification problems and provides three contributions: a GP-based statistical verification framework for stochastic systems, a metric to estimate the accuracy of the framework's predictions, and active sampling algorithms specifically developed for closed-loop statistical verification.  Given a small set of stochastic simulations, the first two contributions will predict the probability of requirement satisfaction at all possible parametric uncertainties and compute the confidence in those predictions.  The third contribution exploits this prediction confidence to select subsequent simulations and minimize prediction error.  Three case studies are provided to demonstrate the effectiveness of the approach to maximize accuracy given a constraint on the number of simulations.

%%%%%%%%%%%%%%
\section{Problem Description}\label{s:prob}
Consider the following nonlinear system
\begin{equation}
	\dot{\vec{x}}(t) = f(\vec{x}(t),\vec{u}(t),\vtheta,\vec{w}(t)),
\end{equation}
where $\vec{x}(t) \in \mathbb{R}^n$ is the state vector, $\vec{u}(t) \in \mathbb{R}^m$ is the control input vector, $\vtheta \in \mathbb{R}^p$ represents the parametric uncertainties, and $\vec{w}(t) \in \mathbb{R}^n$ is the stochastic noise.  A candidate controller generates control inputs according to some predetermined function $\vec{u}(t) = g(\vec{x}(t))$.  Given this controller, the resulting closed-loop dynamics reduce to 
\begin{equation}\label{eq:system}
	\dot{\vec{x}}(t) = f_{cl}(\vec{x}(t),\vtheta,\vec{w}(t)),
\end{equation}
and the trajectory $\Phi(\vec{x}(t)|\vec{x}_0,\vtheta)$ is a function of nominal initial state $\vec{x}_0$, parametric uncertainties $\vtheta$, and stochastic noise $\vec{w}(t)$.  We assume the nominal initial state $\vec{x}_0$ is known and fixed.  

The parametric uncertainties $\vtheta$ can arise from a variety of sources.  For instance, these parameters may include system properties such as vehicle mass and inertia that will typically vary across different operational scenarios.  The main impact of the parametric uncertainties is as variable initial conditions that determine the evolution of the subsequent state trajectory $\Phi(\vec{x}(t)|\vec{x}_0,\vtheta)$.  These parametric uncertainties may not be known during real-world execution of the system, but are assumed to fall within some bounded, compact set $\Theta$ that is known in advance.
{\parindent=0pt 
	\begin{assumption}\label{as:compact}
	The set of all possible parametric uncertainties $\vtheta \in \Theta$ is a known, compact set $\Theta \in \mathbb{R}^p$.  
\end{assumption}}
Assumption~\ref{as:compact} ensures the system has known feasible bounds on the uncertain conditions for the verification procedure to examine.  Most physical systems will naturally have bounds on the set of allowable operating conditions.  For instance, aircraft will typically fly within a maximum take-off weight set by structural limitations and a minimum (empty) weight.  

The major difference between \cref{eq:system} and the deterministic systems found in earlier problems\cite{Quindlen18_GNC,Quindlen17a_Arxiv,Kozarev16_HSCC} is the addition of stochastic noise $\vec{w}(t)$ in the dynamics.  This noise term encapsulates any source of randomness, such as process and/or measurement noise and ultimately breaks the previous works' fundamental assumption of deterministic (noise-free) dynamics.  The randomness will cause multiple trajectories starting from the same $\vec{x}_0$ and $\vtheta$ to obtain different results.  The noise could have a large effect on the satisfaction of the performance requirements at a given $\vec{x}_0$ and $\vtheta$.

%%%%%%%%%%%%%%%%%%%%%%%%%%
\subsection{Measurements of Trajectory Performance}
A trajectory must meet certain performance requirements in order for it to be considered ``satisfactory.''  As before\cite{Quindlen17a_Arxiv}, these requirements are supplied by relevant certification authorities and may include a wide range of potential specifications with varying complexity.  Commonly, the requirements can be written as signal temporal logic (STL)\cite{Maler04_FORMATS} specifications, which also offers a convenient method to quantify the satisfaction of the requirements.  Central to STL is the computation of a scalar robustness degree $\rho^{\varphi} \in \mathbb{R}$ that quantifies the minimum robustness of the trajectory with respect to the requirement specification $\varphi$.   Therefore, the satisfaction of the requirements along the entire trajectory is captured with a single scalar variable.  Building upon that concept, this work assumes the robustness of a trajectory $\Phi(\vec{x}(t)|\vec{x}_0,\vtheta)$ to a given performance requirement is measured with a single measurement variable $y(\vtheta)$, where $y(\theta) = \rho^{\varphi}$ in problems with STL specifications.  Other applications will specify the performance requirements and quantify their satisfaction through different methods, as will be seen in the task allocation example from Section \ref{s:cbba}.
{\parindent=0pt 
\begin{assumption}\label{as:measurement}
	The simulation model provides a single, scalar output $y(\vtheta) \in \mathbb{R}$ that quantifies the minimum robustness of trajectory $\Phi(\vec{x}(t)|\vec{x}_0,\vtheta)$ to a specified performance requirement.  The sign of $y(\vtheta)$ indicates satisfaction of the requirements, where $y > 0$ signifies ``satisfactory'' performance and $y\leq0$ signifies ``unsatisfactory'' performance. 
\end{assumption}}
Since the closed-loop trajectory is a function of nominal (fixed) $\vec{x}_0$ and $\vtheta$, we write the trajectory robustness measurements as an explicit function of $\vtheta$ to emphasize the effect of the parametric uncertainties upon the robustness of the trajectory.  As $\vtheta$ changes, so will the satisfaction of the performance requirement.

%Different applications will produce robustness measurements $y(\vtheta)$ through various mean.  For example, signal temporal logic (STL)\cite{Maler04_FORMATS} is a widely-used mathematical framework for modeling the requirements with temporal logic specifications.  Central to this framework is the computation of a scalar robustness degree $\rho^{\varphi} \in \mathbb{R}$ that quantifies the minimum robustness of a trajectory with respect to a set of temporal logic specifications $\varphi$ (the requirement) and is an obvious potential source for measurement $y(\vtheta)$.  Other applications will quantify the satisfaction of the performance requirements through different methods like realized reward or score functions, as will be seen in the task allocation example from Section \ref{s:cbba}.
 
%%%%%%%%%%%%%%%%%%%%%%%%
\subsection{Probabilistic Satisfaction of Requirements}
Although a closed-loop trajectory and its corresponding robustness measurement are functions of parametric uncertainties, stochasticity will cause multiple trajectories at the same $\vtheta$ setting to obtain different $y(\vtheta)$.  If a large number of simulations are performed at this setting, then the corresponding $y(\vtheta)$ measurements will produce a continuous distribution of possible robustness values.  The underlying true distribution of $y(\vtheta)$ may take a variety of different forms, but this work addresses the case with a Gaussian distribution. %\XX{is th is a major limitation? seems like we might need to discuss - I could see reviewers getting hung up on this pt}
{\parindent=0pt 
\begin{assumption}\label{as:gaussian}
	The distribution of robustness measurements $y(\vtheta)$ at every $\vtheta$ follows a Gaussian distribution $y(\vtheta) \sim \mathcal{N}(\bar{y}(\vtheta),\epsilon_y^2)$ with spatially-varying mean $\bar{y}(\vtheta)$ and constant standard deviation $\epsilon_y$.  
\end{assumption}}
%\noindent Although this assumption does restrict the class of problems, the paper lays the groundwork for a more complex framework able to handle non-Gaussian distributions.  Upcoming work will relax Assumption \ref{as:gaussian} and present extensions to this paper's baseline approach to consider spatially-varying $\epsilon_y$ and non-Gaussian distributions.  In these extensions, the implementation details for the prediction model in Section \ref{s:predictionModel} will change, but the rest of the approach in Sections \ref{s:modeling} and \ref{s:active} remains the same and is only slightly modified to match the new implementation changes. %%%%1st attempt
%\noindent This assumption does restrict the class of problems, but the fundamental concepts and approach are the same for more complex distributions.  Assumption \ref{as:gaussian} simplifies the implementation details in Section \ref{s:modeling} and enables a much clearer presentation of the contributions in Sections \ref{s:confidence} and \ref{s:active}.  Upcoming work will relax Assumption \ref{as:gaussian} and present extensions to consider spatially-varying $\epsilon_y$ and non-Gaussian distributions, but the approach remains the same and only the implementation details change. %%%%2nd revision
\noindent While this assumption does restrict the class of problems, it simplifies the implementation details to improve reader clarity and better highlight the contributions of the approach.  This paper is intended to introduce the statistical verification framework and discuss the challenges and contributions on clearer, and still relevant, verification problems.  Upcoming extensions will relax Assumption \ref{as:gaussian} to consider spatially-varying $\epsilon_y$\cite{Lazaro11_ICML} and non-Gaussian distributions\cite{Seiferth17_ACC}, but also require significantly more complex and sensitive statistical inference techniques to model those distributions.  The implementation details will change in these extensions, but the fundamental approach remains the same. %%%%%3rd revision

From the verification perspective, where each trajectory either satisfies the requirement or does not, the Gaussian distribution of $y(\vtheta)$ creates a Bernoulli distribution for the likelihood an arbitrary trajectory at $\vtheta$ will satisfy the requirement.
{\parindent=0pt 
\begin{definition}\label{def:true_CDF}
	The satisfaction probability function $p_{sat}(\vtheta) \in [0,1]$ defines the probability an arbitrary trajectory initialized with $\vtheta$ will satisfy the performance requirement.
\end{definition}}
This probability of satisfaction is the cumulative distribution of positive $y(\vtheta)$ measurements,
\begin{equation}\label{eq:true_CDF}
	p_{sat}(\vtheta) = \mathbb{P}(y(\vtheta)>0) = \frac{1}{2} + \frac{1}{2} \text{erf }\bigg(\frac{\bar{y}(\vtheta)}{\sqrt{2\epsilon_y^2}}\bigg).
\end{equation}
%While $p_{sat}(\vtheta)$ defines the probabilistic satisfaction of the requirements, the challenge for stochastic verification is $p_{sat}(\vtheta)$ will generally be unknown in advance.  Instead, the statistical verification goal is to predict $p_{sat}(\vtheta)$ for all possible conditions in $\Theta$.  Likewise, the verification process must also contend with computational costs.  It may seem straightforward to simply saturate $\Theta$ with simulations, but the complexity of the simulation model \jhmargin{impart a computational budget on statistical verification.}{doesn't make much sense}  This work models the computational budget as a feasible upper bound $N_{total}$ on the number of simulation trajectories.  The complete statistical verification problem is summarized below.
%\begin{problem}\label{prob:objective}
%	Given a stochastic closed-loop simulation model \cref{eq:system}, compute the estimated satisfaction probability function $\hat{p}_{sat}(\vtheta)$ given a limited number of simulation trajectories, $N_{total}$.
%\end{problem}
While $p_{sat}(\vtheta)$ defines the probabilistic satisfaction of the requirements, the challenge for stochastic verification is $p_{sat}(\vtheta)$ will generally be unknown in advance.  Instead, the statistical verification goal is to compute an estimated $\hat{p}_{sat}(\vtheta)$ for all possible conditions in $\Theta$ and minimize the difference between the predictions and $p_{sat}(\vtheta)$.  

Simultaneously, statistical verification must also contend with computational costs.  In most applications, external constraints restrict the computational budget allocated to the statistical verification process.  For instance, verification typically relies upon high-fidelity models of complex systems and subsystems that drive up the computational time for a single trajectory.  It is infeasible to simply saturate $\Theta$ with multiple simulations at every $\vtheta$ because each simulation trajectory requires a certain amount of time and computational resources.   Regardless of the source, each application has some feasible limit on the amount of time, resources, or money allocated to the statistical verification process.  This work models the feasible limit as an upper bound $N_{total}$ on the number of simulation trajectories.  %The complete statistical verification problem is summarized below.
%\begin{problem}\label{prob:objective}
%	Given a set of simulation trajectories $\mathcal{L}$, compute the estimated satisfaction probability function $\hat{p}_{sat}(\vtheta)$ to solve
%	\begin{equation}
%	\begin{aligned}
%		\text{minimize } & | p_{sat}(\vtheta_i) - \hat{p}_{sat}(\vtheta_i)| \ \ \forall \vtheta_i \in \Theta \\
%			\text{s.t.} & \ |\mathcal{L}| \leq N_{total}
%	\end{aligned}
%	\end{equation}
%\end{problem}

%%%%%%%%%%%%%%
\section{Statistical Verification Framework}\label{s:modeling}
Given the verification objective, there are multiple possible approaches to compute $\hat{p}_{sat}(\vtheta)$ from a finite set of simulation-based observations.  The most basic form is to treat each trajectory as a Bernoulli trial with binary evaluations \{``satisfactory'', ``unsatisfactory''\} from the sign of $y(\vtheta)$.  While this approach can address all types of systems, including those that do not meet Assumptions \ref{as:measurement} and \ref{as:gaussian}, a binary-based approach requires multiple simulations at every $\vtheta$ to construct a binomial distribution.  Each repetition counts against the sampling budget $|\mathcal{L}| \leq N_{total}$ and means one few simulation at another location.  Although the cost of each simulation is less of a problem for applications with simple simulation models, the cost of training a prediction model given a large number of simulations at each location is non-negligible.  Therefore, the large number of simulations required for practical predictions with binomial distributions restricts the tractability of those approaches.

Instead, this work entirely avoids the need for expensive binomial distributions in the relevant class of systems through the direct use of $y(\vtheta)$ and its Gaussian probability density function (PDF).  The main idea is to exploit a single stochastic  trajectory at each training location and its noisy measurement $y(\vtheta)$ to infer the underlying PDF, which defines the cumulative distribution $p_{sat}(\vtheta)$.  Although various computational methods are possible\cite{Bishop07_PRML,Tipping01_JMLR}, this problem is particularly well-suited to Gaussian process regression\cite{Rasmussen06} due to the Gaussian distribution of $y(\vtheta)$.

%%%%%%%%%%%%%%%%%%
\subsection{GP-based Prediction Model}\label{s:predictionModel}
The Gaussian process regression model follows a similar format to the earlier work in verification of deterministic systems\cite{Quindlen17a_Arxiv}.  A finite collection of $N$ total simulation trajectories forms a training dataset $\mathcal{L} = \{\mathcal{D},\vec{y}\}$ consisting of their parameter settings $\mathcal{D} = \{\vtheta_1,\vtheta_2,\hdots,\vtheta_N\}$ and corresponding robustness values $\vec{y} = [y(\vtheta_1),\hdots,y(\vtheta_N)]^T$.  Unlike the deterministic approach, these stochastic measurements require the introduction of a Gaussian likelihood model, where the GP does not model $y(\vtheta)$ directly, but infers the latent mean $\bar{y}(\vtheta)$.

%The training process constructs the GP regression model from the information provided by the training dataset $\mathcal{L}$.  More information is found in \cite{Rasmussen06}, but fundamentally the training procedure uses Bayesian inference to place a posterior probability distribution on latent mean $\bar{\vec{y}}$ given $\mathcal{L}$.  Assuming a zero-mean prior $\mathbb{P}(\bar{\vec{y}}|\mathcal{D},\psi) = \mathcal{N}(\bar{\vec{y}}|\vec{0},\vec{K})$ and likelihood model $\mathbb{P}(\vec{y}|\bar{\vec{y}}, \vartheta) = \mathcal{N}(\vec{y}|\bar{\vec{y}},\epsilon_y^2\vec{I})$, the posterior predictive distribution at an arbitrary location $\vtheta_*$ is $\mathbb{P}(\bar{\vec{y}}(\vtheta_*)|\mathcal{L},\vtheta_*,\psi,\vartheta) = \mathcal{N}(\mu(\vtheta_*),\Sigma(\vtheta_*))$.  The posterior predictive mean $\mu(\vtheta_*)$ and covariance $\Sigma(\vtheta_*)$ are given by
The training process constructs the GP regression model from the information provided by the training dataset $\mathcal{L}$.  More details are found in \cite{Rasmussen06}, but fundamentally the training procedure uses Bayesian inference to place a posterior probability distribution on latent mean $\bar{\vec{y}}$ given $\mathcal{L}$.  Assuming a zero-mean prior $\mathcal{N}(\bar{\vec{y}}|\vec{0},\vec{K})$ and likelihood model $\mathcal{N}(\vec{y}|\bar{\vec{y}},\epsilon_y^2\vec{I})$, the posterior predictive distribution at an arbitrary location $\vtheta_*$ follows a Gaussian distribution $\mathbb{P}(\bar{\vec{y}}(\vtheta_*)|\mathcal{L},\vtheta_*) = \mathcal{N}(\mu(\vtheta_*),\Sigma(\vtheta_*))$.  The posterior predictive mean $\mu(\vtheta_*)$ and covariance $\Sigma(\vtheta_*)$ are given by
\begin{equation}\label{eq:GP}
\begin{aligned}
	\mu(\vtheta_*) & = \vec{K}_*^T \big( \vec{K} + \epsilon_y^2\vec{I}\big)^{-1} \vec{y} \\
	\Sigma(\vtheta_*) & =  \kappa(\vtheta_*,\vtheta_*) - \vec{K}_*^T \big( \vec{K} + \epsilon_y^2\vec{I}\big)^{-1} \vec{K}_* \ ,
\end{aligned}
\end{equation}
where $\kappa(\vtheta_i,\vtheta_j)$ is the scalar kernel function, $\vec{K}_*$ is the $N \times 1$ vector of $\kappa(\vtheta_*,\vtheta_i) \ \forall i=1:N$ and $\vec{K}$ is the $N \times N$ matrix for $\kappa(\vtheta_i,\vtheta_j) \ \forall i,j=1:N$.  Different choices are possible, but this work uses the common squared exponential kernel with automatic relevance determination (SE-ARD)\cite{Rasmussen06} for the kernel function.  

The choice in GP hyperparameters may drastically change the predictive distribution, even with the same training set $\mathcal{L}$, since these hyperparameters control the kernel function $\kappa$ and the likelihood model.  Unfortunately, the optimal choice of hyperparameters that perfectly replicates $\bar{y}(\vtheta)$ will not be known in advance and must be estimated online.  This work uses maximum likelihood estimation (MLE)\cite{Rasmussen06} to optimize the kernel hyperparameters given only the information provided by $\mathcal{L}$.  The use of the SE-ARD kernel also enables the MLE procedure to adjust the sensitivity of the kernel to each element of $\vtheta$ as $\bar{y}(\vtheta)$ may have a higher sensitivity to certain elements than others.  
Additionally, the likelihood model contains a hyperparameter to estimate standard deviation $\epsilon_y$ as this term may also be completely unknown.  If $\epsilon_y$ is unknown, then an estimate $\hat{\epsilon}_y$ may be computed in the same MLE optimization process as the kernel hyperparameters or separately by repeated sampling at a training locations in $\mathcal{D}$.

%%%%%%%%%%%%%%%%%%%%%%%
\subsubsection{Expected Probability of Satisfaction}
Although $\mu(\vtheta_*)$ and $\Sigma(\vtheta_*)$ define the predictive PDF for $\bar{y}(\vtheta_*)$, \cref{eq:GP} only completes half the goal.  Definition \ref{def:true_CDF} requires the cumulative distribution function (CDF) for $y(\vtheta_*)$, not the PDF.  The CDF in \cref{eq:true_CDF} defined the true $p_{sat}(\vtheta)$, but this computation requires perfect knowledge of $\bar{y}(\vtheta)$ and $\epsilon_y$.  Instead, the predicted satisfaction probability function $\hat{p}_{sat}(\vtheta)$ marginalizes the CDF over the posterior predictive distribution of $\bar{y}(\vtheta)$,
\begin{equation}\label{eq:CDF}
\begin{aligned}
	\hat{p}_{sat}(\vtheta_*) & = \mathbb{E}_{\bar{y}(\vtheta_*)} \Big[ \mathbb{P}\big(y(\vtheta_*) > 0|\bar{y}(\vtheta_*)\big) \Big] \\
		& = \frac{1}{2} +  \frac{1}{2} \text{erf }\bigg(\frac{\mu(\vtheta)}{\sqrt{2( \Sigma(\vtheta_*) + \epsilon_y^2)}}\bigg).
\end{aligned}
\end{equation}
%This result provides the stochastic verification framework's solution to Problem \ref{prob:objective}.  %The choice in hyperparameters will also affect the expected probability of satisfaction.  

%%%%%%%%%%%%%%%%%%%%%%%
\subsection{Prediction Confidence}\label{s:confidence}
While \cref{eq:CDF} computes the expected probability of requirement satisfaction, $\hat{p}_{sat}(\vtheta)$ will likely fail to perfectly model $p_{sat}(\vtheta)$.  Therefore, the statistical verification framework must not only provide $\hat{p}_{sat}(\vtheta)$, but also indicate where its confidence in the accuracy of these predictions is low.  Unsurprisingly, the true prediction error $\tilde{p}_{sat}(\vtheta) = p_{sat}(\vtheta) - \hat{p}_{sat}(\vtheta)$ is unknown, but offline validation methods using external validation datasets\cite{Bishop07_PRML} can estimate $\tilde{p}_{sat}(\vtheta)$ by comparing the predictions against the validation set's known, true values.  However, these external validation methods are wasteful as they generally require the removal of valuable training data for the independent validation set.

For comparison, probabilistic inequalities provide theoretically-justified bounds on $\tilde{p}_{sat}(\vtheta)$ without the use of validation sets.  In particular, Chebyshev's inequality bounds the prediction error by the variance of the CDF,
\begin{equation}\label{eq:chebyshev}
	\mathbb{P}(|\tilde{p}_{sat}(\vtheta)| \geq a) \leq \frac{\mathbb{V}_{\bar{y}(\vtheta_*)} \Big[ \mathbb{P}\big(y(\vtheta_*) > 0|\bar{y}(\vtheta_*)\big) \Big] }{a^2} \ ,
\end{equation}
where $a>0$.  Lower CDF variance will translate into lower probabilistic bounds on $\tilde{p}_{sat}(\vtheta)$ and thus higher confidence in the accuracy of $\hat{p}_{sat}(\vtheta)$.  The primary challenge with \cref{eq:chebyshev} is the variance lacks an analytical closed-form solution.  Fortunately, the variance can be approximated using a 1st or 2nd order Taylor series expansion of the nonlinear random function\cite{Ang07}.  The 1st order approximation is
\begin{equation}\label{eq:variance}
	\mathbb{V}_{\bar{y}(\vtheta_*)} \big[ \mathbb{P}(y(\vtheta_*) > 0|\bar{y}(\vtheta_*)) \big] \approx \frac{1}{2\pi \epsilon_y^2} e^{-\mu(\vtheta_*)^2/\epsilon_y^2} \ \Sigma(\vtheta_*),
\end{equation}
which also happens to conservatively upper bound the 2nd order approximation.  For simplicity, we subsequently refer to the approximate CDF variance in \cref{eq:variance} as $V(\vtheta|\mathcal{L})$.

Despite the convenience of \cref{eq:variance}, the accuracy of an approximation of a nonlinear random function is limited.  This inaccuracy makes it inadvisable to blindly substitute \cref{eq:variance} into \cref{eq:chebyshev} without careful consideration.  However, the approximation in \cref{eq:variance} does still provide a perfect metric to identify regions of $\Theta$ where the confidence in $\hat{p}_{sat}(\vtheta)$ is low and provide intuition on the sensitivity of $\tilde{p}_{sat}(\vtheta)$.  The power of \cref{eq:variance} is its value as a computationally-efficient, online validation tool for signifying prediction confidence.  For instance, approximate variance is highest in regions with small $|\mu(\vtheta)|$ \emph{and} large $\Sigma(\vtheta)$, meaning the confidence in the predictions should be quite low.  The next section will exploit the approximate CDF variance to derive a closed-loop verification process that seeks to minimize prediction errors.

%%%%%%%%%%%%%%
\section{Closed-Loop Statistical Verification}\label{s:active}
As discussed in Section \ref{s:prob}, statistical verification will have some feasible limit on the amount of time or computational resources allocated to the overall process. This limit is approximated as a cap on the number of simulation trajectories, $|\mathcal{L}| \leq N_{total}$.  Given this restriction on the size of training dataset $\mathcal{L}$, the ideal scenario would only perform informative simulations and carefully select all $\vtheta \in \mathcal{D}$ to minimize the prediction error over all $\vtheta \in \Theta$.  However, the information-maximizing training dataset is not apparent until after all the trajectories have been obtained.  To address that issue, this work applies active learning\cite{Settles12} to iteratively select informative $\vtheta$ settings for future simulations and minimize the expected prediction error.  To emphasize the iterative, feedback-based nature of the active learning procedure, we label the process \emph{closed-loop statistical verification}.  

Active learning describes a wide variety of different procedures\cite{Settles12,Kremer14_DMKD,Chen16_CDC,Desautels14_JMLR,Gotovos13_IJCAI,Zhang16_AAAI}, each with their own definition for the ``best'' sample to run next.  Most of these procedures focus on a particular aspect of the Gaussian PDF.  For instance, procedures focused on the PDF mean\cite{Kremer14_DMKD,Quindlen18_GNC} favor points with $\bar{y}(\vtheta)$ near zero, meaning the best location is $\overline{\vtheta} = \text{argmin}|\mu(\vtheta)|$.  Although originally intended for binary classification with support vector machines, such an approach does correctly label points with low $|\mu(\vtheta)|$ as informative since the CDF variance \cref{eq:variance} is high there.  Likewise, PDF variance-based approaches and extensions\cite{Chen16_CDC,Desautels14_JMLR,Gotovos13_IJCAI,Zhang16_AAAI} are significantly more common for GP methods and aim to reduce the PDF variance $\Sigma(\vtheta)$.  These approaches favor points with high variance, $\overline{\vtheta} = \text{argmax } \Sigma(\vtheta)$, which also correctly emphasizes points with comparatively high CDF variance since $\Sigma(\vtheta)$ will be large.  Although they both correctly emphasize certain aspects, neither of those two approaches explicitly minimizes the posterior CDF variance.

%%%%%%%%%%%%%%%%%
\subsection{Reduction in CDF Variance}
Earlier work in deterministic closed-loop verification\cite{Quindlen17a_Arxiv} specifically developed new sample selection metrics to maximize prediction confidence.  Even though the implementation details have changed, this work utilizes the same underlying motivation and attempts to minimize the approximate CDF variance \cref{eq:variance} in order to maximize the confidence in $\hat{p}_{sat}(\vtheta)$.  The ideal selection criterion would minimize the maximum posterior CDF variance $V(\vtheta|\mathcal{L}^+)$ after the new sample data at $\overline{\vtheta}$ was added, but this requires the posterior training set $\mathcal{L}^+ = \mathcal{L} \cup \{\overline{\vtheta},y(\overline{\vtheta})\}$ to be known apriori.  Prior knowledge of $\mathcal{L}^+$ is an impossible proposition since $y(\overline{\vtheta})$ cannot be known before a simulation has actually been performed there.  Even if the expected posterior set $\hat{\mathcal{L}}^+ = \mathcal{L} \cup \{\overline{\vtheta},\mu(\overline{\vtheta})\}$ replaces infeasible $\mathcal{L}$, the high computational cost of retraining the GP at every prospective sample location, nominally an $\mathcal{O}(N^3)$ operation and $\mathcal{O}(N^2)$ at best, renders the approach computationally intractable.  

A more computationally tractable approach maximizes the local improvement in posterior CDF variance, effectively selecting the $\vtheta$ setting which will experience the greatest reduction in CDF variance if a simulation is performed at that location.  The local change in CDF variance is labeled by $\tilde{V}(\vtheta|\mathcal{L}) = V(\vtheta|\mathcal{L}) - V(\vtheta|\hat{\mathcal{L}}^+)$.  Although the change $\tilde{V}(\vtheta|\mathcal{L})$ requires expected posterior information $\hat{\mathcal{L}}^+$, it does not require the GP model to be retrained, which was the source of the previous computational intractability.  The local posterior CDF variance at location $\vtheta_*$ after a measurement there can be written purely in terms of the current information.  After the Woodbury matrix identity, the expected posterior covariance $\Sigma(\vtheta_*)^+$ reduces to
\begin{equation}
	\Sigma(\vtheta_*)^+ = \Sigma(\vtheta_*)\Big(1 - \frac{\Sigma(\vtheta_*)}{\Sigma(\vtheta_*) + \epsilon_y^2} \Big)
\end{equation}
while mean $\mu(\vtheta_*)^+ = \mu(\vtheta_*)$.  Ultimately, the local change in CDF variance is given by
\begin{equation}\label{eq:varRed}
	\tilde{V}(\vtheta_*|\mathcal{L}) = \frac{1}{2\pi \epsilon_y^2} e^{-\mu(\vtheta_*)^{2}/\epsilon_y^2} \ \Sigma(\vtheta_*) \bigg( \frac{\Sigma(\vtheta_*)}{\Sigma(\vtheta_*) + \epsilon_y^2}\bigg).
\end{equation}
Note that the CDF variance reduction criterion in \cref{eq:varRed} favors points with low $|\mu(\vtheta)|$ and high $\Sigma(\vtheta)$, just as in \cref{eq:variance}.

%%%%%%%%%%%%%%%
\subsection{Sampling Algorithms}
The section criterion from \cref{eq:varRed} forms the basis of the closed-loop stochastic verification framework.  Since the actual set $\Theta$ is infinite, the framework constructs a high resolution lattice $\Theta_d$ to replicate $\Theta$ and then selects all training locations from $\Theta_d$ rather than $\Theta$ directly $(\mathcal{D} \subset \Theta_d)$.  Set $\mathcal{U} = \Theta_d \setminus \mathcal{D}$ contains all remaining points not in training set $\mathcal{L}$ available for future simulations.  This paper presents two versions of the closed-loop verification framework.

%%%%%%%%%%%%%%%%%
\subsubsection{Sequential Sampling}
The most straightforward implementation of closed-loop verification is sequential sampling, described in Algorithm \ref{alg:sequential}.  The process starts with an initial training set $\mathcal{L}$ of passively-selected samples, usually obtained through random sampling or another open-loop design of experiments approach.  This initial training set of size $|\mathcal{L}| = N_0$ produces the initial GP so active sampling can be performed.  The sequential procedure then selects one sample from $\mathcal{U}$ according to \cref{eq:varRed} before it performs one simulation at the selected $\overline{\vtheta}$ vector, obtains $y(\overline{\vtheta})$, and updates the GP prediction model.  This iterative procedure repeats until the number of remaining samples $T = N_{total} - N_0$ has been reached.

\begin{algorithm}[t]
\caption{\strut Sequential closed-loop stochastic verification}
\label{alg:sequential}
\begin{algorithmic}[1]
	\STATE \textbf{Input:} initial training set $\mathcal{L} = \{\mathcal{D},\vec{y}\}$, available sample locations $\mathcal{U}$, max \# of additional samples $T$
	\STATE \textbf{Initialize:} train GP regression model
	\FOR{i=1:T}
	\STATE{Select $\overline{\theta} = \underset{\vtheta \in \mathcal{U}}{\text{ argmax }} \tilde{V}(\vtheta|\mathcal{L})$}
	\STATE{Perform simulation at $\overline{\vtheta}$, obtain measurement $y(\overline{\vtheta})$}
	\STATE{Add $\{\overline{\vtheta},y(\overline{\vtheta})\}$ to training set $\mathcal{L}$, remove $\overline{\vtheta}$ from $\mathcal{U}$}
	\STATE{Retrain GP model with updated $\mathcal{L}$}
	\ENDFOR
	\STATE \textbf{Return:} expected $\hat{p}_{sat}(\vtheta)$ and variance $V\big(\vtheta|\mathcal{L}\big)$
\end{algorithmic}
\end{algorithm}

%%%%%%%%%%%%%%%%
\subsubsection{Batch Sampling}
In comparison to Algorithm \ref{alg:sequential}, batch sampling\cite{Settles12} offers further computational efficiency by selecting multiple samples between retraining steps.  Assuming the same limit $N_{total}$, batch selection reduces the number of GP retraining processes and can also fully exploit any parallel computing capabilities of the simulation environment.  Batch approaches select $M$ samples at once and perform their simulations in parallel.  While this provides computational savings, it also introduces the possibility of redundant samples if the batch sample set $\mathcal{S}$ does not possess adequate diversity.  

First proposed for deterministic closed-loop verification in \cite{Quindlen17a_Arxiv}, determinantal point processes (DPPs)\cite{Kulesza11_ICML} present efficient probabilistic methods for encouraging diversity within $\mathcal{S}$ without severe computational overhead.  The batch framework first converts $\tilde{V}(\vtheta|\mathcal{L})$ into a probability distribution $\mathbb{P}_V(\vtheta) = \tilde{V}(\vtheta|\mathcal{L})/Z_V$, where $Z_V$ is the normalization constant.  A relatively small number of $M_T$ samples are obtained from $\Theta_d$ according to $\mathbb{P}_V(\vtheta)$ and used to construct a DPP.  This paper uses $M_T = 1000$, but these are simply samples of $\vtheta$ locations from the distribution and not actual simulations are performed.  The DPP randomly selects $M$ locations from $\mathcal{U}$ for $\mathcal{S}$ based upon a modified version of $\mathbb{P}_V$ that penalizes similarities in $\mathcal{S}$ and subsequently spreads the datapoints out across high-valued regions in $\Theta_d$ with significantly less redundancy.  Algorithm \ref{alg:batch} details the batch closed-loop verification framework.  Rather than $T$ number of additional simulations, Algorithm \ref{alg:batch} operates in $T$ batches of $M$ simulations, assuming $M\cdot T \leq N_{total} - N_0$.

\begin{algorithm}[t]
\caption{\strut Batch closed-loop stochastic verification framework using determinantal point processes}
\label{alg:batch}
\begin{algorithmic}[1]
	\STATE \textbf{Input:} initial training set $\mathcal{L} = \{\mathcal{D},\vec{y}\}$, available sample locations $\mathcal{U}$, \# of iterations $T$, batch size $M$
	\STATE \textbf{Initialize:} train GP regression model
	\FOR{i=1:T}
	\STATE \textbf{Initialize:} $\mathcal{S} = \emptyset$
	\STATE{Transform $\tilde{V}(\vtheta|\mathcal{L})$ into distribution $\mathbb{P}_V(\vtheta)$}
	\STATE{Form k-DPP from $M_T$ random samples of $\mathbb{P}_V(\vtheta)$}
	\STATE{Generate $M$ random samples from DPP, add to $\mathcal{S}$}
	\STATE{Run simulation $\forall \vtheta\in\mathcal{S}$, obtain measurements $\vec{y}_{\mathcal{S}}$}
	\STATE{Add $\{\mathcal{S},\vec{y}_{\mathcal{S}}\}$ to training set $\mathcal{L}$, remove $\mathcal{S}$ from $\mathcal{U}$}
	\STATE{Retrain GP model with updated $\mathcal{L}$}
	\ENDFOR
	\STATE \textbf{Return:} expected $\hat{p}_{sat}(\vtheta)$ and variance $V\big(\vtheta|\mathcal{L}\big)$
\end{algorithmic}
\end{algorithm}

%%%%%%%%%%%%%%
\section{Examples}\label{s:results}
This section examines statistical verification applied to three stochastic systems and demonstrates the improved performance of Algorithms \ref{alg:sequential} and \ref{alg:batch} over existing techniques.  

%%%%%%%%%%%%%%%%%%%%%%%%%%%%%%%%%%%%%
\subsection{Example 1: Model Reference Adaptive Control System}\label{s:clmrac}
The first example considers a stochastic version of the concurrent learning model reference adaptive control (CL-MRAC) system from earlier deterministic work\cite{Quindlen18_GNC,Quindlen17a_Arxiv}.  The CL-MRAC example examines a second order linear system with two uncertain parameters $\vtheta = [\theta_1,\theta_2]^T$.  The adaptive control system estimates these parameters online and attempts to track a desirable reference trajectory.  The nonlinearities associated with adaptation result in a highly nonlinear closed-loop system.  

The verification goal is to predict the probability the actual trajectory will remain within 1 unit of the reference trajectory at all times between $0\leq t \leq 40$ seconds, given in STL format as
\begin{equation}
	\varphi = \Box_{[0,40]} (1 - |e_1[t]| \geq 0)
\end{equation}
where tracking error $e_1(t)$ is the difference between actual position $x_1(t)$ and reference position $x_{m_1}(t)$.  The STL operator $\Box_{[0,40]}$ refers to ``for all times between $t=0$ and $t=40$ seconds.''  Measurement $y(\vtheta)$ is the STL robustness degree $\rho^{\varphi}$ for the trajectory initialized with uncertainty setting $\vtheta$.  The sampling lattice $\Theta_d$ covers $\theta_1: [-10,10]$ and $\theta_2:[-10,10]$ with a grid of 40,401 points.  In this example, the open-loop dynamics are subject to additive process noise with Gaussian distribution $\vec{w}(t) = \mathcal{N}(\vec{0},\vec{I})$.  The underlying true distribution was obtained by repeated sampling at each of the points in $\Theta_d$.  To avoid any potential issues, Gaussian distributions were fit to the raw data and the variance was averaged across $\Theta_d$ to return $\epsilon_y = 0.0372$.  %Extensions with $\vtheta$-varying $\epsilon_y$ are discussed later.  

\begin{figure}[!]
	\centering
	{\includegraphics[width=.89\columnwidth]{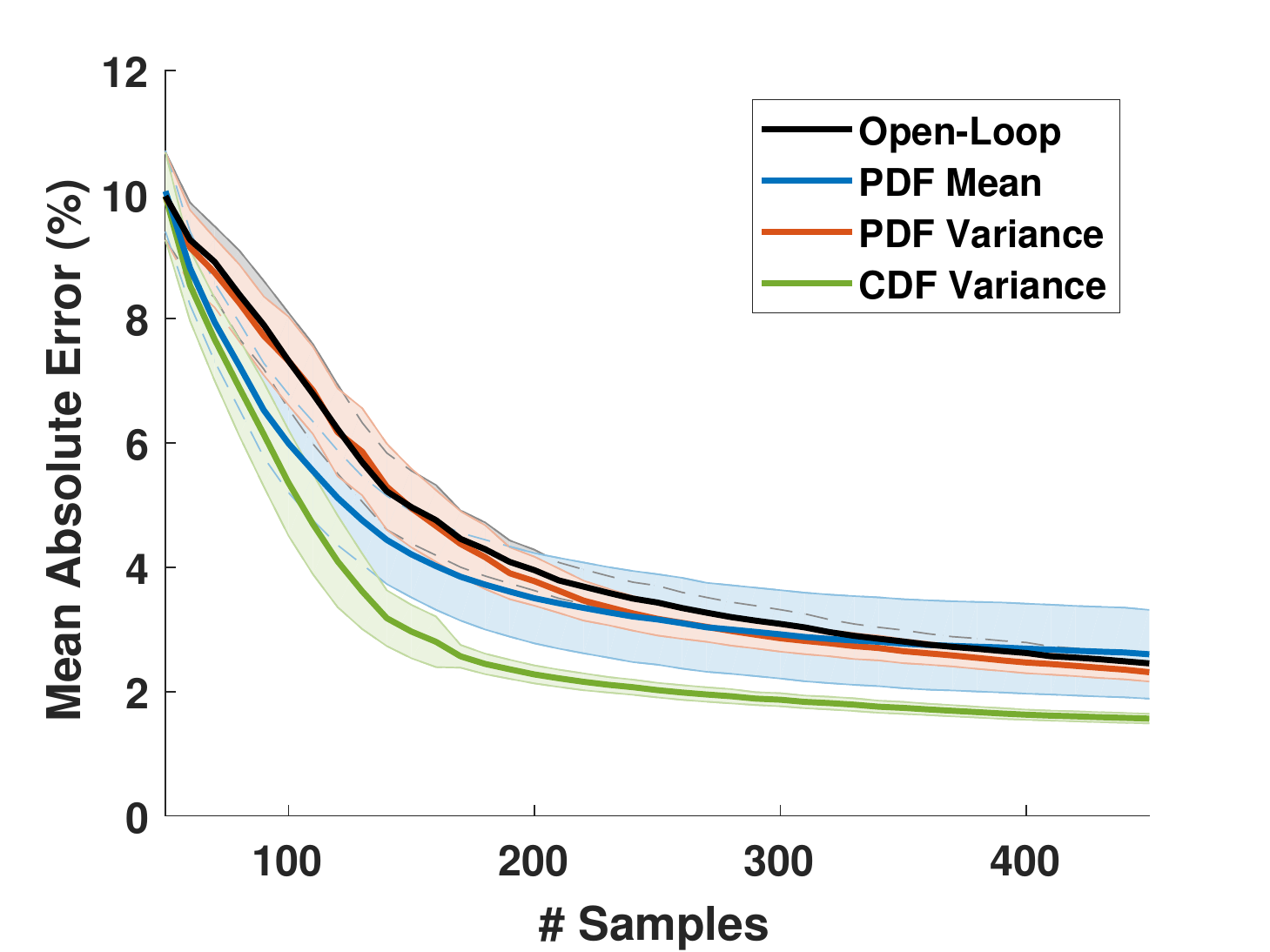}}
	\caption{(Example 1) Comparison of mean absolute error (MAE) convergence for the four different sampling strategies.  The standard deviation intervals around the mean (solid lines) are given by the 0.5$\sigma$ bound.}
 		\label{f:mrac1}
	\vspace{-0.in}
\vspace{0.2in}
	\centering
	{\includegraphics[width=.89\columnwidth]{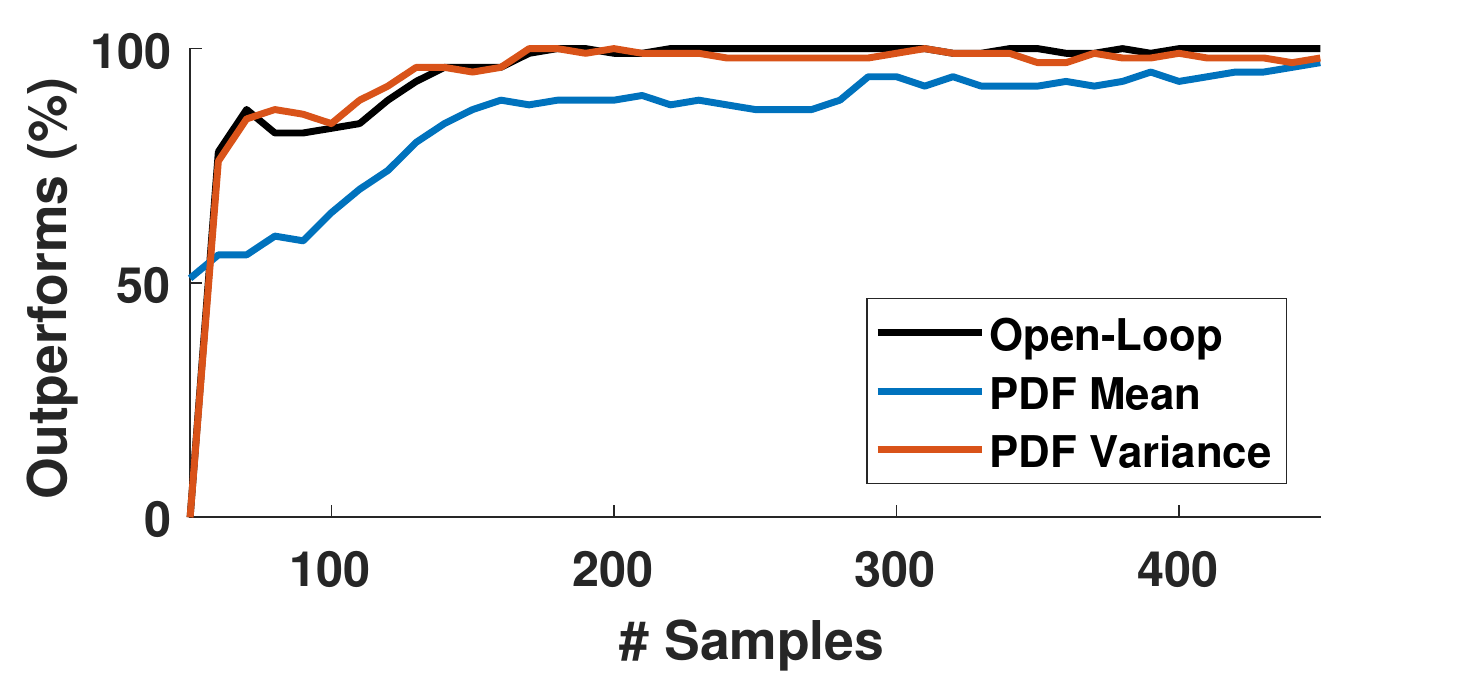}}
	\caption{(Example 1) Ratio of cases where Algorithm \ref{alg:batch} directly outperforms or matches the MAE of the indicated strategies.}
 		\label{f:mrac2}
	\vspace{-0.1in}
\end{figure}

\fig{f:mrac1} compares the performance of Algorithm \ref{alg:batch} against similar procedures using the existing selection metrics discussed in Section \ref{s:active} as well as an open-loop, random sampling procedure.  These procedures all start with an initial training set of 50 simulations and select batches of $M=10$ points until a total of 450 simulations has been reached.  Neither the kernel hyperparameters nor $\epsilon_y$ are assumed to be known so the procedures estimate these online using maximum likelihood estimation.  In order to fairly compare each procedure, the algorithms all start from the same 100 randomly-chosen initial training sets with the same random seed.  At the conclusion of the process, Algorithm \ref{alg:batch} demonstrates a 29\%, 31\%, and 35\% improvement in average mean absolute error (MAE) over the PDF mean, PDF variance, and random sampling approaches.  Additionally, the performance of the algorithms in each of the 100 test cases can be directly compared since they all start with the same $\mathcal{L}$ and random seed.  At this level, \fig{f:mrac2} illustrates that Algorithm \ref{alg:batch} will either eventually match or outperform the existing sampling strategies nearly 100\% of the time.  Although the exact numbers will change for different distributions, these results highlight the value of the selection criteria from \cref{eq:varRed} to further reduce prediction error given a fixed number of samples.

\begin{figure}[!]
	\centering
	{\includegraphics[width=.89\columnwidth]{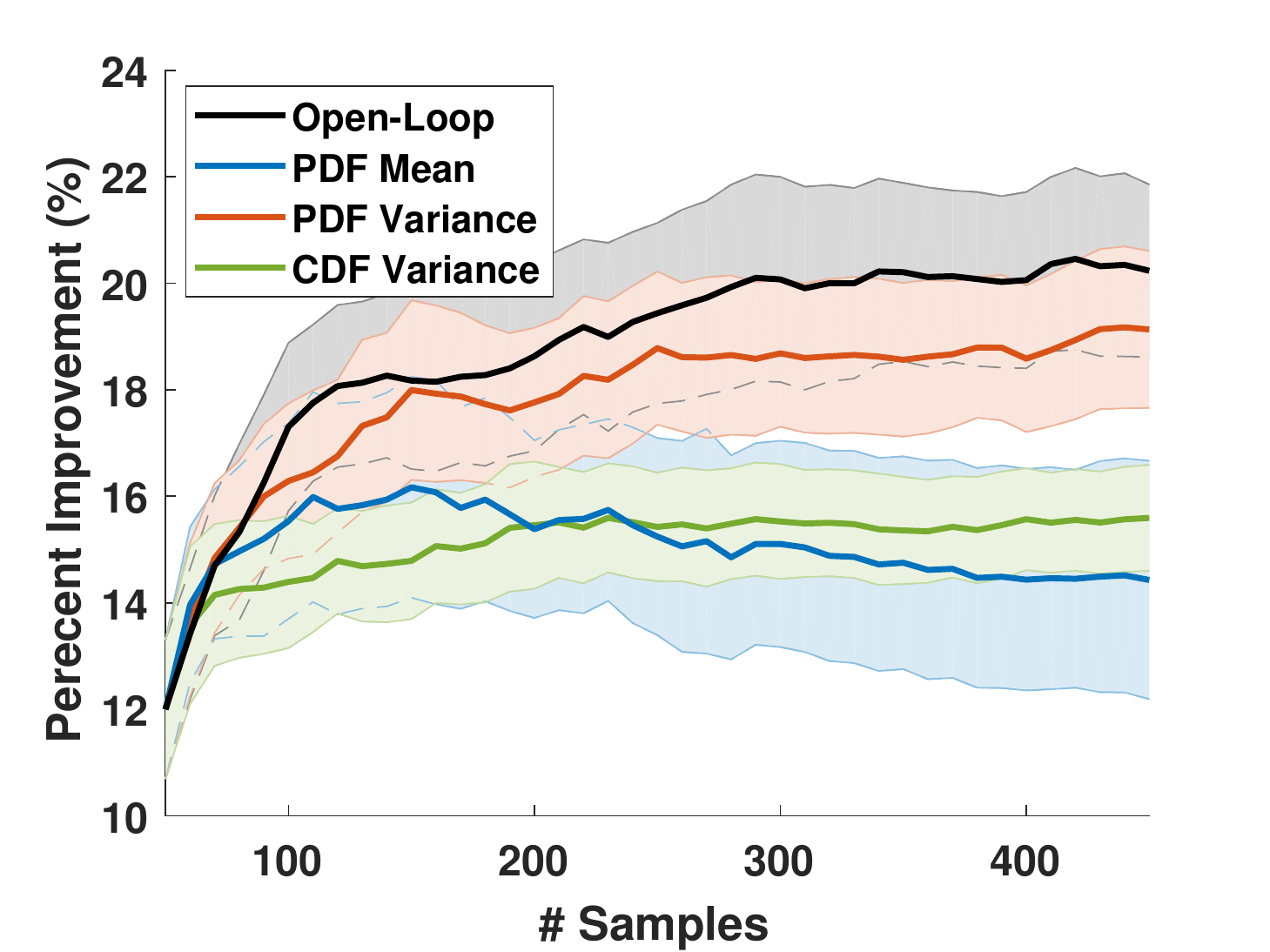}}
	\caption{(Example 1) Reduction in MAE after all points with the top 5\% of CDF variance are removed.}
 		\label{f:mrac3}
	\vspace{-0.1in}
\end{figure}

\fig{f:mrac3} demonstrates the utility of CDF variance to identify regions of large prediction errors, regardless of the particular sampling strategy.  As discussed in Section \ref{s:confidence}, the true CDF variance is unavailable, but the approximate CDF variance \cref{eq:variance} still provides a meaningful metric to compare prediction confidence across $\Theta_d$.  One of the best uses of \cref{eq:variance} is to rank points in $\Theta_d$ according to their CDF variance in order to identify which predictions to trust the least.  \fig{f:mrac3} displays the reduction in prediction error when the points with the top 5\% of CDF variance are removed.  The 12-22\% improvement in MAE proves that the CDF variance did indeed correctly identify and remove regions with large $\tilde{p}_{sat}(\vtheta)$.  When the top 10\% is removed, the improvement jumps to 40\%.  

%%%%%%%%%%%%%%%%%%%%%%%%%%%%%%%%
\subsection{Example 2: Robust Multi-Agent Task Allocation}\label{s:cbba}
The second example addresses the robust multi-agent task allocation problem from \cite{Quindlen17_GNC} with added stochasticity.  The task allocation problem attempts to assign a team of four UAVs to complete fire surveillance tasks that will take longer or shorter depending on the wind speed $\theta_1$ and direction $\theta_2$.  Task durations are also corrupted by zero-mean Gaussian noise.  Unforeseen time delays will compound and may potentially lead the UAVs to miss the completion of tasks within their assigned window, thus lowering the overall realized mission score.  The verification goal is to determine whether the team of UAV agents will sufficiently complete the ordered tasking and achieve a minimum mission score at different wind settings.  Sampling grid $\Theta_d$ spans the set of feasible wind conditions $\theta_1: [0\degrees, 359\degrees], \theta_2: [0, 40]$ km/hr with 16,641 possible trajectory settings.

\begin{figure}[!]
	\centering
	{\includegraphics[width=.89\columnwidth]{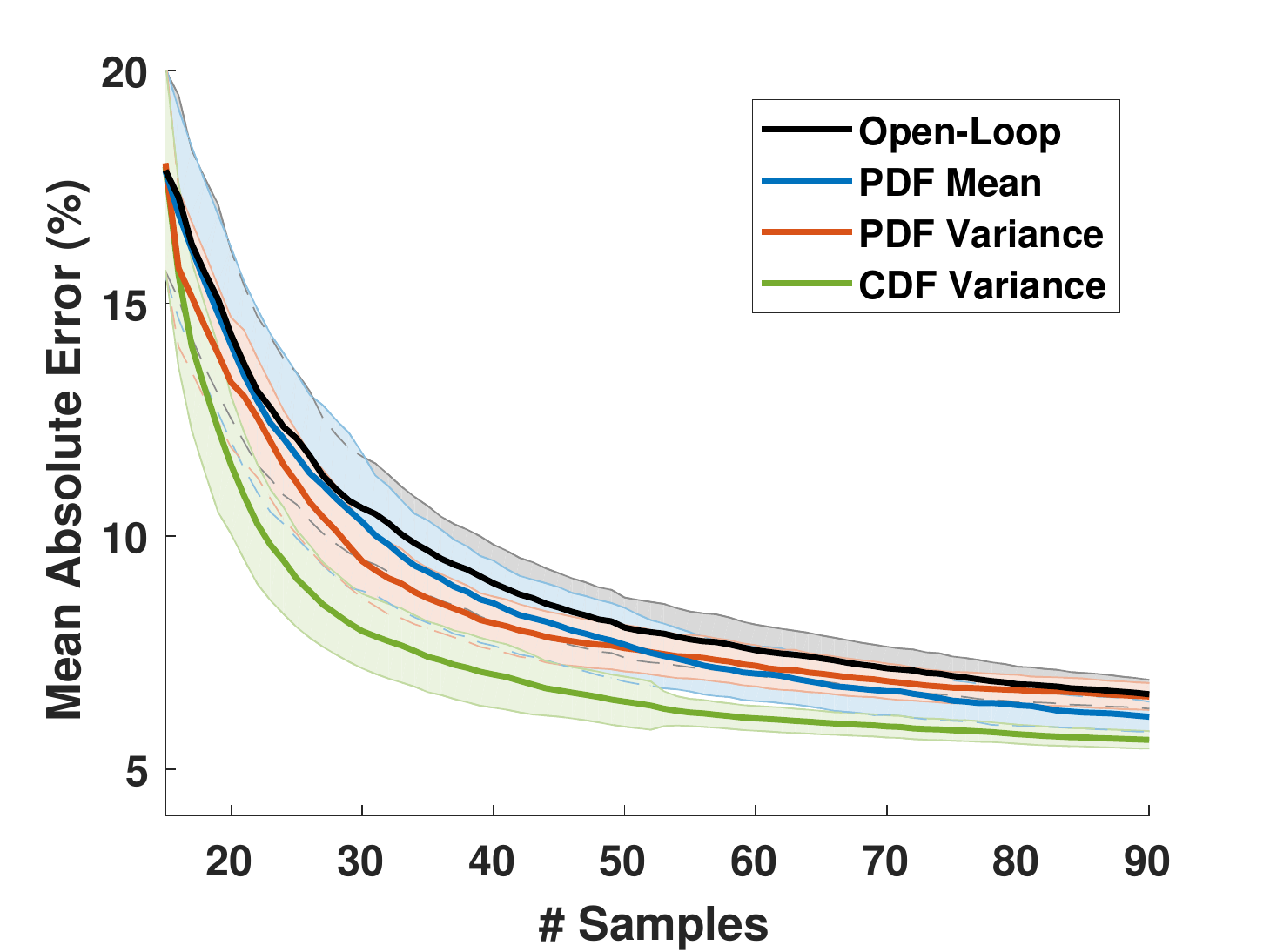}}
	\caption{(Example 2) Comparison of MAE convergence for the four different sampling strategies over 250 randomly selected initializations.}
 		\label{f:cbba1}
	\vspace{-0.1in}
\vspace{0.2in}
	\centering
	{\includegraphics[width=.89\columnwidth]{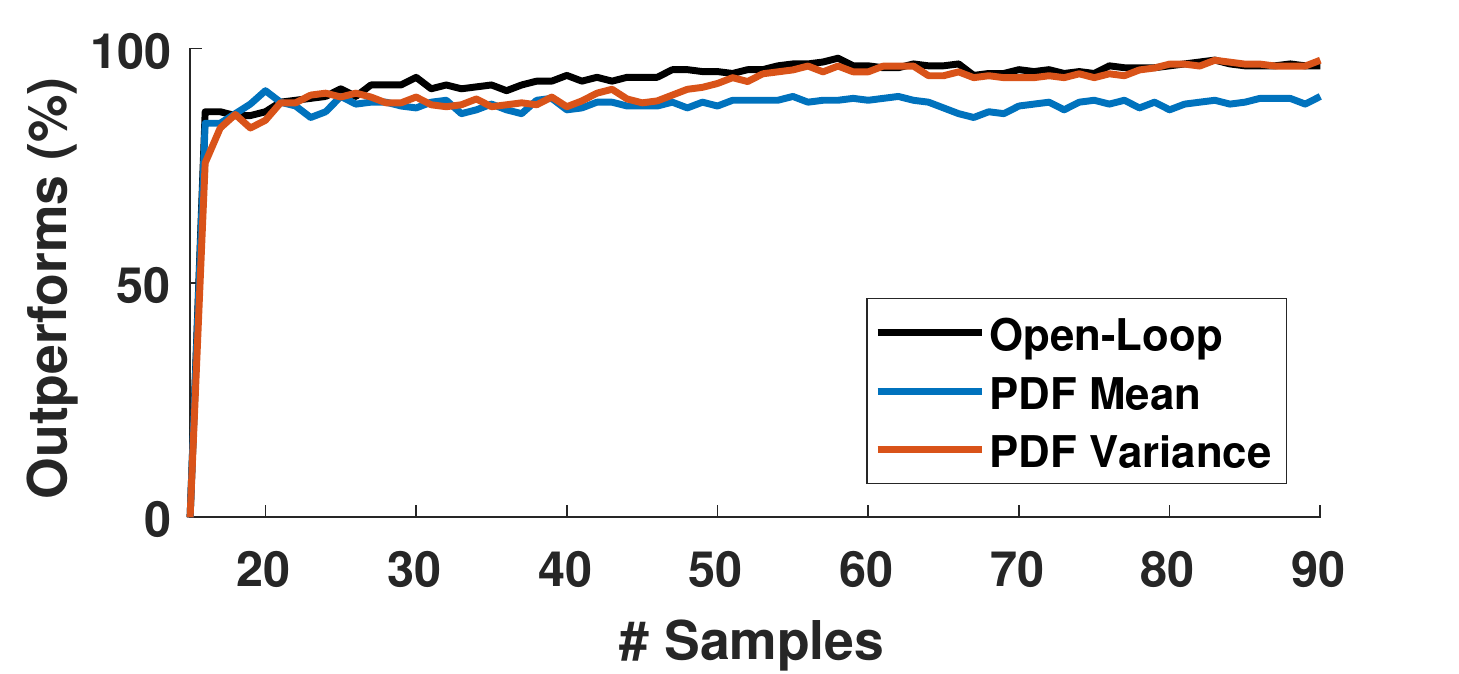}}
	\caption{(Example 2) Ratio of cases where Algorithm \ref{alg:sequential} directly outperforms or matches the MAE of the indicated strategies.}
 		\label{f:cbba2}
	\vspace{-0.1in}
\end{figure}

Figures \ref{f:cbba1} and \ref{f:cbba2} compare the performance of Algorithm \ref{alg:sequential} against the competing sampling strategies for 250 randomly-initialized test cases.  %This example still optimizes the kernel hyperparameters with MLE, but now numerical approximates true $\epsilon_y$ through repeated samples at a few locations.  
Ultimately, the MAE performance is consistent with the last example.  Algorithm \ref{alg:sequential} demonstrates a 10-20\% improvement in average MAE over the existing strategies and either matches or exceeds the MAE of the competing approaches when directly compared against one another.  

\begin{figure}[!]
	\centering
	{\includegraphics[width=.89\columnwidth]{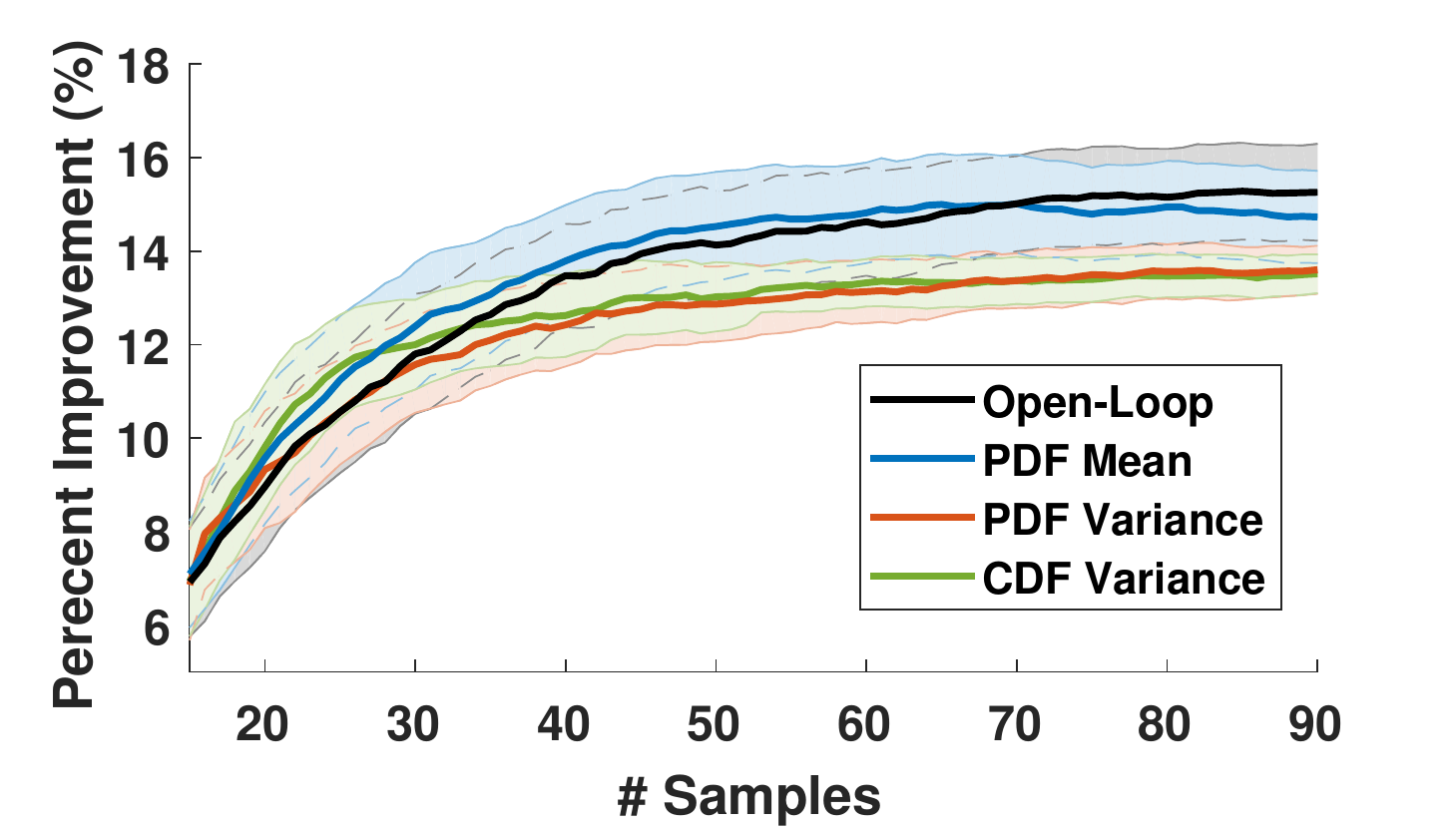}}
	\caption{(Example 2) Reduction in MAE after all points with the top 5\% of CDF variance are removed.}
 		\label{f:cbba3}
	\vspace{-0.1in}
\end{figure}

\fig{f:cbba3} demonstrates a similar reduction in MAE as witnessed in \fig{f:mrac3}.  Once the points with the top 5\% of CDF variance are removed, the MAE of the predictions in the remaining 95\% of the data reduces by up to 12-16\%.  If the top 10\% are removed, the average reduction reaches at least 28\% for all the sampling strategies.  This supports the conclusion that CDF variance identifies points with low prediction confidence where $\tilde{p}_{sat}(\vtheta)$ may be large.  %It's a probability-based metric so it will miss points with large errors, but \cref{eq:variance} is the only computational tool for any form of online validation without external validation datasets.  

%%%%%%%%%%%%%%%%%%%%%%%%
\subsection{Example 3: Lateral-Directional Autopilot}\label{s:autopilot}
Lastly, the third example considers the altitude-hold requirement of a lateral-directional autopilot\cite{Elliott16_S5}.  In this problem, a Dryden wind turbulence model\cite{MIL1797} augments the original nonlinear aircraft dynamics model to introduce stochasticity into the closed-loop dynamics.  The performance requirement expects the autopilot to maintain altitude $x(t)$ within a set threshold of the initial altitude when the autopilot is engaged.  This work explores a relaxed version of the original requirement\cite{Elliott16_S5} in which the aircraft must remain within a 55 foot window around the initial altitude $x(0)$,
\begin{equation}
	\varphi = \Box_{[0,50]} (55 - |x[t] - x[0]| \geq 0).
\end{equation} 
The performance measurement is the STL robustness degree $\rho^{\varphi}$.  This example tests the satisfaction of the requirement against initial Euler angles for roll $\theta_1: [-60\degrees,60\degrees]$, pitch $\theta_2: [4\degrees,19\degrees]$, and yaw $\theta_3: [75\degrees,145\degrees]$ and longitudinal inertia $\theta_4: [5430,8430] (kg \cdot m^2)$, with a constant reference heading of 112\degrees.  The 4D grid $\Theta_d$ consists of 937,692 possible parameter settings for the simulations.

\begin{figure}[!]
	\centering
	{\includegraphics[width=.89\columnwidth]{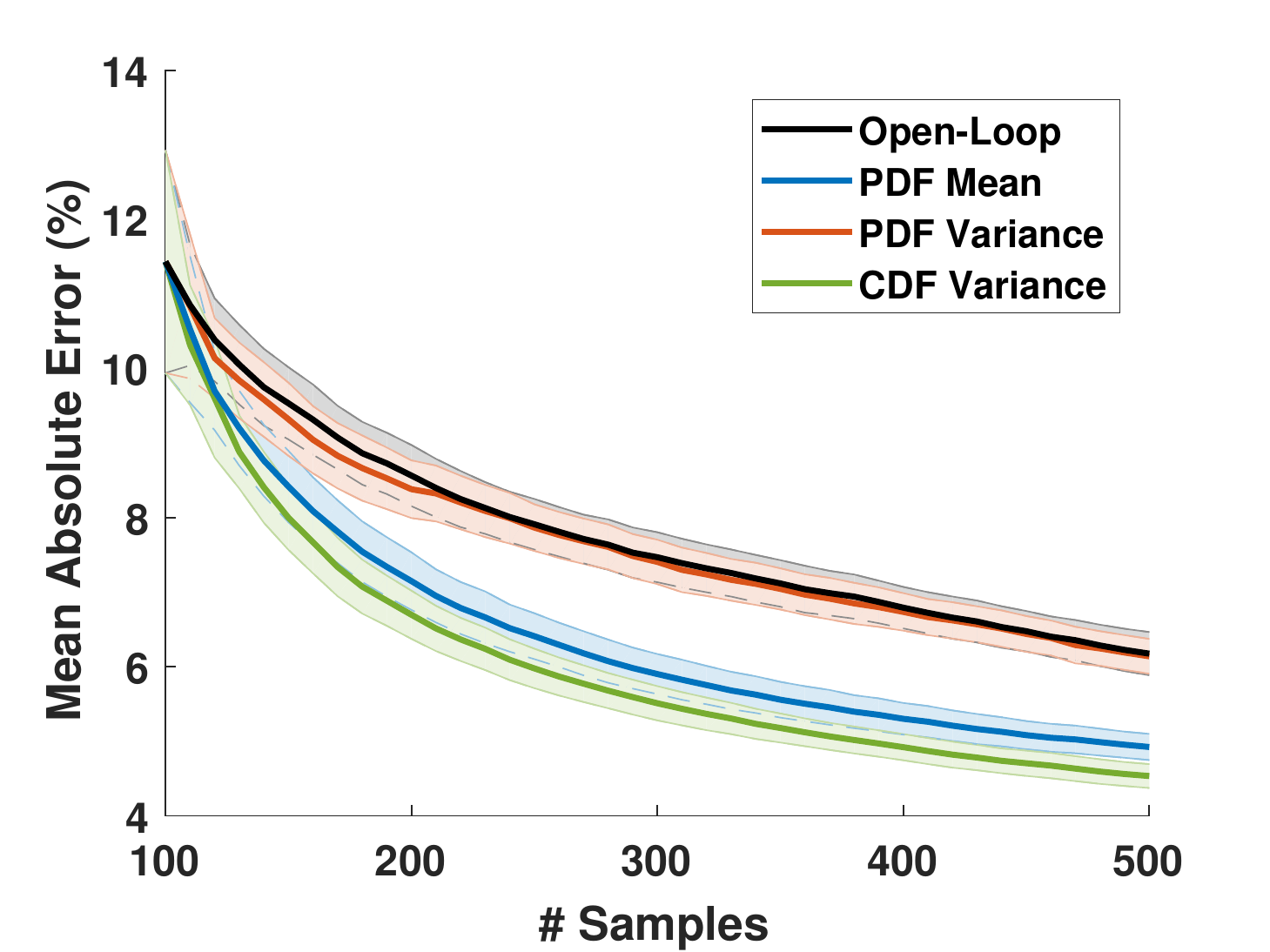}}
	\caption{(Example 3) Comparison of MAE convergence for the four different sampling strategies over 120 random initializations.}
 		\label{f:lockheed1}
	\vspace{-0.1in}
%\end{figure}
%
%\begin{figure}[!]
\vspace{0.2in}
	\centering
	{\includegraphics[width=.89\columnwidth]{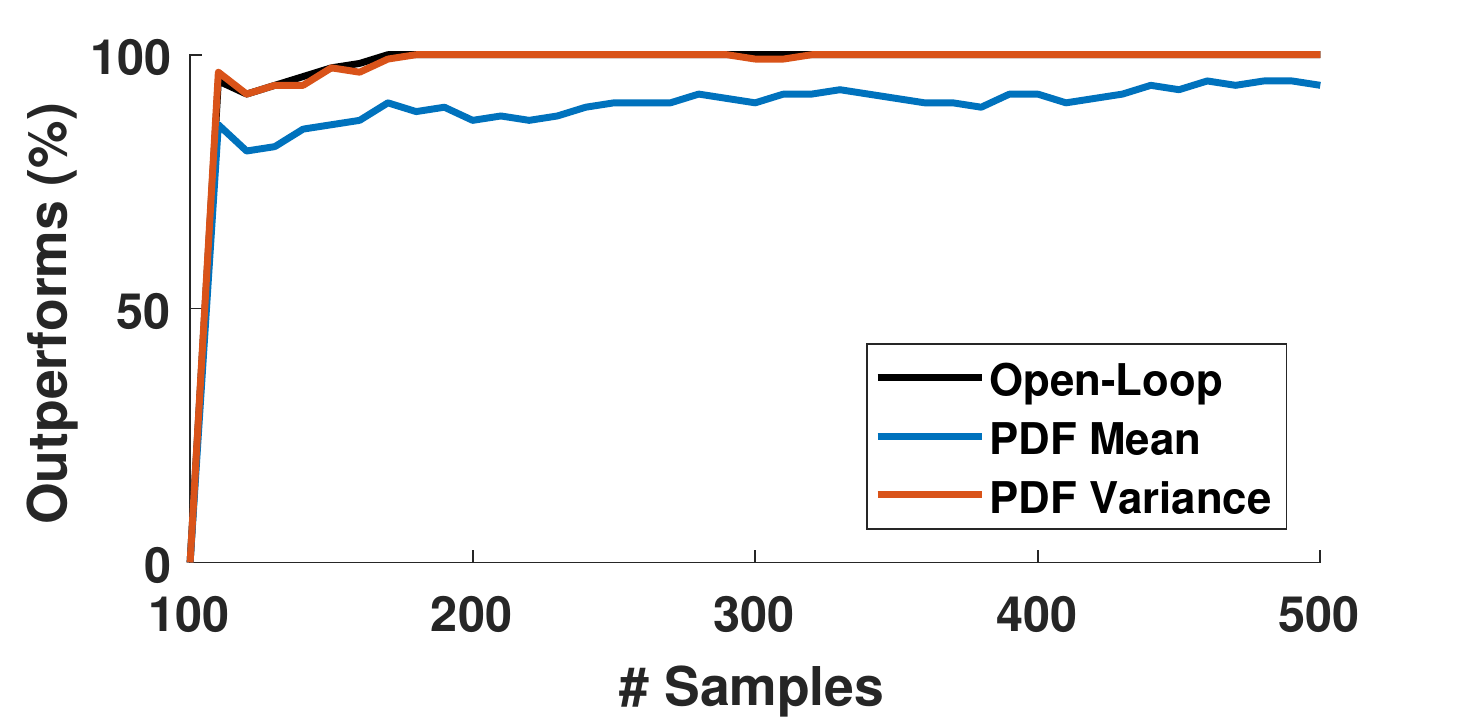}}
	\caption{(Example 3) Ratio of cases where Algorithm \ref{alg:batch} directly outperforms or matches the MAE of the indicated strategies.}
 		\label{f:lockheed2}
	\vspace{-0.1in}
\end{figure}

Figures \ref{f:lockheed1} and \ref{f:lockheed2} illustrate the MAE performance of the four different sampling strategies given a batch size of $M=10$.  The equivalent of Figures \ref{f:mrac3} and \ref{f:cbba3} is skipped but displays the same exact behavior.  This example begins with 100 simulations in the initial training set $\mathcal{L}$ and compares the performance over 120 random initializations.  As before, Algorithm \ref{alg:batch} outperforms the PDF variance-focused and random sampling procedures with a 27\% reduction in average MAE.  Unlike the previous two examples, Algorithm \ref{alg:batch} only slightly edges out the competing mean-focused strategy.  This new result is due to the comparatively low $\epsilon_y$ for noisy $y(\vtheta)$ from the low-altitude turbulence model.  The resulting tighter distribution only experiences changes in $p_{sat}(\vtheta)$ near $\bar{y}(\vtheta) = 0$, which favors the mean-focused selection criteria.  If the turbulence is increased then the distribution widens dramatically and the performance of the PDF mean-focused metric degrades significantly like the previous two examples.  These case studies all highlight the improved performance of the novel CDF variance algorithms compared to the existing sampling strategies.  Just as important, Algorithms \ref{alg:sequential} and \ref{alg:batch} will remain the best-performing sampling strategy even as the distribution changes, whereas the performance of the other strategies will vary according to changes in $\epsilon_y$.

%%%%%%%%%%%%%%
\section{Conclusion}\label{s:conclude}
This work introduced machine learning methods for simulation-based statistical verification of stochastic nonlinear systems.  In particular, the paper presented a GP-based statistical verification framework to efficiently predict the probability of requirement satisfaction over the full space of possible parametric uncertainties given a limited amount of simulation data.  Additionally, Section \ref{s:confidence} developed new criteria based upon the variance of the cumulative distribution function to qualify confidence in the predictions.  This metric provides a simple validation tool for online identification of regions where the prediction confidence is low.  Section \ref{s:active} builds upon this metric and introduces sequential and batch sampling algorithms for efficient closed-loop verification.  These new verification procedures demonstrate up to a 35\% improvement in prediction error over competing approaches in the three examples in Section \ref{s:results}.  The examples also serve to highlight the utility of the CDF variance to correctly identify low-confidence regions in the parameter space without external validation datasets.  

While the paper only considers Gaussian distributions for the performance measurements, this work lays the foundation for more complex statistical verification frameworks capable of handing a wider range of distributions.  Upcoming work will adapt the CDF variance metric and closed-loop verification procedures to recent developments in modeling of spatially-varying standard deviations\cite{Lazaro11_ICML} and non-Gaussian distributions\cite{Seiferth17_ACC}.  Additionally, this work can be extended to high-dimensional GP representations\cite{Kandasamy15_ICML}.  While the implementation considers Gaussian distributions, the fundamental concepts have broader utility.

\section*{Acknowledgments} This work is supported by the Office of Naval Research and Air Force Office of Scientific Research under grants ONR MURI N00014-11-1-0688 and  AFOSR FA9550-15-1-0146 .

\balance

%%%%%%%%%%%%%%%%%%%%%%%%%%%%%%%%%%%%%%%%%%%%%%%%%%%%%%%%%%%%%%%%%%%%%%%%%%%%%%%%
\bibliographystyle{IEEEtran}
\bibliography{acc_2018,BIB_all/ACL_Publications,BIB_all/ACL_all}

%%%%%%%%%%%%%%%
%\section{Appendix}
%\input{sec_appendix}

\end{document}